\documentclass[twocolumn]{IEEEtran}
\usepackage{amsmath}
\usepackage{amsfonts}
\usepackage{amssymb}
\usepackage{graphicx}
\def\BibTeX{{\rm B\kern-.05em{\sc i\kern-.025em b}\kern-.08em
    T\kern-.1667em\lower.7ex\hbox{E}\kern-.125emX}}
\begin{document}

\title{The Bloch Oscillating Transistor }
\author{Heikki Sepp\"{a} and Juha Hassel\thanks{Version 2 submitted to APL March 16th
2001}\thanks{The Authors are with the VTT Automation, Measurement Technology,
VTT Technical Research center of Finland, P.O. Box 02044 VTT, Finland
(telephone and e-mail of the corresponding author +358-9-456-6771,
Heikki.Sepp\"{a}@vtt.fi)}}
\maketitle
\begin{abstract}
We introduce a new mesoscopic transistor, which consists of a superconducting
island connected to superconducting and normal electrodes via two mesoscopic
tunnel junctions. Furthermore, the island is being charged through a resistor.
The interplay between Bloch oscillations, single-electron effects and ohmic
current leads to a device having a high current gain. The operation and
characteristics of the transistor are analyzed with a numerical model.
\end{abstract}


As a consequence of improved understanding of phenomena in mesoscopic tunnel
junctions and development of fabrication technologies a variety of
applications has been suggested and realized during the last two decades. The
Single Electron Transistor (SET) \cite{ave1}\ has established its position as
a basic element of nanotechnology. Recently also the Bloch oscillations
\cite{lik1,kuz1} in small superconducting tunnel junctions have been utilized
\cite{zor1,lot1}. In this paper we introduce a new mesoscopic transistor, the
Bloch Oscillating Transistor (BOT). Whereas the SET is more analogous to the
Field Effect Transistor (FET), the BOT seems to have characteristics similar
to those of a Bipolar Junction Transistor (BJT).

If a mesoscopic Josephson junction is connected to a voltage source via a high
impedance, the junction starts to Bloch oscillate. We analyze here a device
where an extra junction enables a single electron to tunnel into an island
bounded by the Josephson junction, the high impedance, and a single electron
junction. We call such a device a single junction BOT (sj BOT) at its
equivalent circuit is shown in Fig. 1. The junction \#1, characterized by its
capacitance $C_{1}$,\ connects the normal metal base electrode to the
superconducting island via a tunnel resistance $R_{T1}$. The Josephson
junction \#2 having a capacitance $C_{2}$ connects the island to the
superconducting emitter. The Josephson coupling energy $E_{J}=\Phi_{0}%
I_{cs}/2\pi$, where $\Phi_{0}$ is the flux quantum and $I_{cs}\,$the critical
current, is assumed to be close to the charging energy of the island
$E_{c}=e^{2}/2C_{\Sigma}$, where $C_{\Sigma}=C_{1}+C_{2}$.%

%TCIMACRO{\FRAME{ftbpFU}{239pt}{149.5pt}{0pt}{\Qcb{The equivalent circuit of
%the single junction Bloch Oscillating Transistor.}}{}{Figure}%
%maintain-aspect-ratio TRUE;  display "USEDEF";  valid_file "T";  width 239pt;
%height 149.5pt;  depth 0pt;  original-width 273.9375pt;
%original-height 170.5625pt;  cropleft "0";  croptop "1";  cropright "1";
%cropbottom "0";  tempfilename 'HES0AS00.wmf';tempfile-properties "XPR";}}}%
%BeginExpansion
\begin{figure}
[ptb]
\begin{center}
\includegraphics[width=8cm]{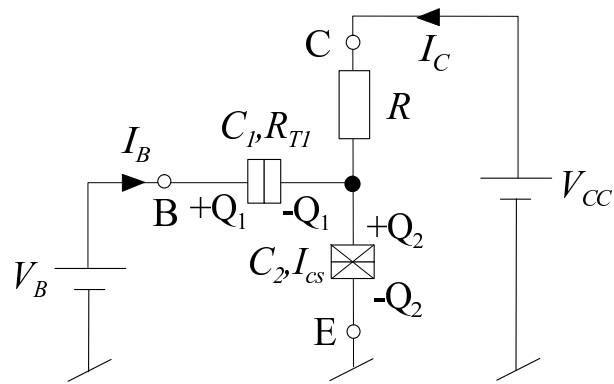}
\caption{The equivalent circuit of the single junction Bloch Oscillating
Transistor.}
\end{center}
\end{figure}
%EndExpansion

We assume here that there is no single electron tunneling through junction
\#2. We further assume that the tunnel resistance $R_{T1}\,$is constant. These
criteria are essentially satisfied if $\left|  V_{CC}\right|  <2\Delta$ and
$\left|  V_{CC}-V_{B}\right|  >\Delta$. Depending on the value of $\Delta$
these criteria are not necessarily satisfied with all voltage values, which
has to be taken into account in the design of a practical device. According to
the theory of Bloch oscillations \cite{lik1}, the quasicharge $Q_{2}%
\,$describes the state of the junction \#2, if the mesoscopic Josephson
junction is sufficiently well isolated. Following the criteria in the theory
of the effect of electromagnetic environment on tunnel junctions \cite{ing1},
we find that the bias resistance $R$ should exceed $R_{q}=h/2e^{2}$.

Since the island is well isolated, the single electron tunneling rates through
junction \#1 can be obtained using the Orthodox Theory with the local rule. In
addition, incoherent Cooper pair tunneling can be neglected. Furthermore, the
second order effects, such as cotunneling and Andre\'{e}v reflections are
neglected. Zero temperature is assumed throughout the analyses.%

%TCIMACRO{\FRAME{ftbpFU}{214.3125pt}{180.125pt}{0pt}{\Qcb{The energy of the
%system as a function of the quasicharge.}}{}{Figure}%
%maintain-aspect-ratio TRUE;  display "USEDEF";  valid_file "T";
%width 214.3125pt;  height 180.125pt;  depth 0pt;  original-width 257.5pt;
%original-height 216.125pt;  cropleft "0";  croptop "1";  cropright "1";
%cropbottom "0";  tempfilename 'HES0AS01.wmf';tempfile-properties "XPR";}}}%
%BeginExpansion
\begin{figure}
[ptb]
\begin{center}
\includegraphics[width=7cm]{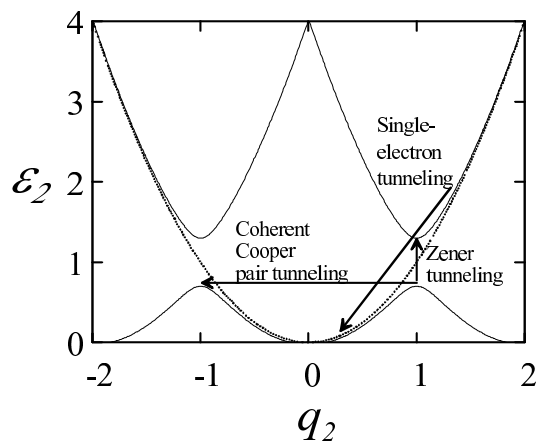}
\caption{The energy of the system as a function of the quasicharge.}%
\end{center}
\end{figure}
%EndExpansion

In this paper the quantities are scaled as follows: The charge is expressed in
units of $e$, the voltage in $e/C_{\Sigma}\,$the current in $e/R_{T1}%
C_{\Sigma}$, the energy in $e^{2}/2C_{\Sigma}$, the resistance in $R_{T1}$,
the capacitance in $C_{\Sigma}$, and the time in $R_{T1}C_{\Sigma} $.

The energy of junction \#2 versus the quasicharge forms $2e$-periodic band
structure. An energy gap, whose magnitude is the Josephson coupling energy
separates the lowest bands\ (the zeroth and the first band). The gap between
bands decreases rapidly as a function of band number, so it is reasonable to
assume that it is zero between the higher bands. The energy is approximated as \cite{gei1}%

\begin{equation}
\varepsilon_{2}(q_{2})=\frac{1}{2}\left[  q_{2}^{2}+\left(  \left|
q_{2}\right|  -2\right)  ^{2}\pm\sqrt{\left(  \left(  \left|  q_{2}\right|
-2\right)  ^{2}-q_{2}^{2}\right)  ^{2}+\varepsilon_{J}^{2}}\right]
\end{equation}
for the two lowest bands, where $q_{2}$ is the quasicharge and $\varepsilon
_{J}$ is the Josephson coupling energy. The minus\ (plus)\ sign corresponds
the zeroth\ (first) band. For higher bands simple parabolic form is assumed,
i.e. $\varepsilon_{2}(q_{2})=q_{2}^{2}$. If $\left|  q_{2}\right|  <e$, the
junction is at the zeroth band, if $e<\left|  q_{2}\right|  <2e$, the junction
is at the first band etc. The energy versus quasicharge $q_{2}$ is illustrated
in Fig. 2.

The charge in the island $q_{I}=q_{2}-q_{1}$, where $q_{1}$ is the charge
across the junction \#1 and $q_{2}$ is the quasicharge of the junction \#2.
The (quasi)charges of individual junctions are obtained from $q_{I}$ with%

\begin{align}
q_{1}  & =\frac{c_{1}\left(  v_{B}-q_{I}/c_{2eff}\right)  }{1+c_{1}/c_{2eff}%
}\\
q_{2}  & =\frac{c_{1}v_{B}+q_{I}}{1+c_{1}/c_{2eff}}.
\end{align}
Here $v_{B}$ is the base voltage, $c_{1}$ is the capacitance of the first
junction and $c_{2eff}\,\,$is the effective capacitance of junction \#2, which
takes into account the band-structure: $c_{2eff}=q_{2}/v_{2}$, where $v_{2}$
is the voltage across junction \#2. The voltage $v_{2}=\frac{1}{2}%
d\varepsilon_{2}/dq_{2}$. The effective capacitance $c_{2eff}$ is a function
of the quasicharge $q_{2}$ (and thus also the time)\ and equals the normal
capacitance everywhere else but near the Brillioun zone boundary, where it diverges.

The change of island charge as a function of time is%

\begin{equation}
\frac{dq_{I}}{d\tau}=i_{C}+\left(  \frac{dq_{I}}{d\tau}\right)  _{SET}+\left(
\frac{dq_{I}}{d\tau}\right)  _{CPT},
\end{equation}
where $i_{C}=$ $\left(  v_{CC}-v_{2}\right)  /r$ is the normalized collector
current, $\left(  dq_{I}/d\tau\right)  _{SET}$ is due to the single electron
tunneling through junction \#1 and $\left(  dq_{I}/d\tau\right)  _{CPT}$
corresponds to the Cooper pair tunneling through junction \#2.%

%TCIMACRO{\FRAME{ftbpFU}{237.375pt}{215.5625pt}{0pt}{\Qcb{The collector current
%as a function of the bias voltage $v_{CC}$ for different base voltages $v_{B}%
%$. The active region, where the coherent Cooper pair current can be controlled
%with single electrons tunneling from the base electrode is dark shaded.}}%
%{}{Figure}{\special{ language "Scientific Word";  type "GRAPHIC";
%maintain-aspect-ratio TRUE;  display "USEDEF";  valid_file "T";
%width 237.375pt;  height 215.5625pt;  depth 0pt;  original-width 279pt;
%original-height 253.25pt;  cropleft "0";  croptop "1";  cropright "1";
%cropbottom "0";  tempfilename 'HES0AS02.wmf';tempfile-properties "XPR";}}}%
%BeginExpansion
\begin{figure}
[ptb]
\begin{center}
\includegraphics[width=7.5cm]
{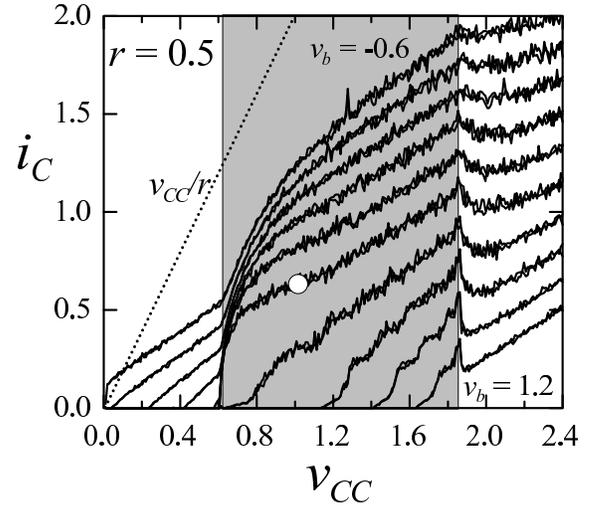}
\caption{The collector current as a function of the bias voltage $v_{CC}$ for
different base voltages $v_{B}$. The active region, where the coherent Cooper
pair current can be controlled with single electrons tunneling from the base
electrode is dark shaded.}%
\end{center}
\end{figure}
%EndExpansion

The change of free energy in a single-electron tunnel event is%

\begin{align}
\Delta f_{1}  & =1+2q_{I}-2\left(  1-c_{1}\right)  v_{b}\\
\Delta f_{2}  & =1-2q_{I}+2\left(  1-c_{1}\right)  v_{b},
\end{align}
where the index $i=1$ ($i=2$) in $\Delta f_{i}$ refers to tunneling from (to)
the island. The tunneling rates are $\gamma_{i}=-(1/2)\Delta f_{i}$, if
$\Delta f_{i}<0$. They are zero otherwise.

As $q_{2}$ approaches the Brillioun zone boundary at $\left|  q_{2}\right|
=e$, the junction \#2 can either tunnel a Cooper pair (move from $+e$ to $-e$
or from $-e$ to $+e$ within the zeroth or first band)\ or move to the higher
(lower) band via Zener-tunneling.\ The latter occurs with probability \cite{gei1}%

\begin{equation}
P_{0\leftrightarrow1}^{Z}=\exp\left[  -\frac{1}{4}\frac{\left(  \varepsilon
_{J}\right)  ^{2}}{\alpha_{t}\left|  i_{C}\right|  }\right]  ,\label{zenprob}%
\end{equation}
where $\alpha_{t}=h/\pi^{2}e^{2}R_{T1}$.

Parameters used in the simulations are $c_{1}=0.1$, $\varepsilon_{J}=0.6$, and
$\alpha_{t}=0.077$. An example of a set of corresponding values in SI units is
$C_{2}=1$ fF, $C_{1}=0.1$ fF, $R_{T1}=34$ k$\Omega$,$\,$and $I_{c}=23$ nA.%

%TCIMACRO{\FRAME{ftbpFU}{229.4375pt}{131.0625pt}{0pt}{\Qcb{ The quasicharge
%$q_{2}\,$versus time at a bias point denoted by a white circle in Fig. 3. The
%Bloch oscillations are interrupted by Zener tunneling, which drives the
%junction to the first energy band.\ A single electron tunneling through
%junction \#1 from the base recovers the system to the zeroth band.}}{}%
%{Figure}{\special{ language "Scientific Word";  type "GRAPHIC";
%maintain-aspect-ratio TRUE;  display "USEDEF";  valid_file "T";
%width 229.4375pt;  height 131.0625pt;  depth 0pt;  original-width 318.25pt;
%original-height 181pt;  cropleft "0";  croptop "1";  cropright "1";
%cropbottom "0";  tempfilename 'HES0AS03.wmf';tempfile-properties "XPR";}}}%
%BeginExpansion
\begin{figure}
[ptb]
\begin{center}
\includegraphics[width=7.5cm]{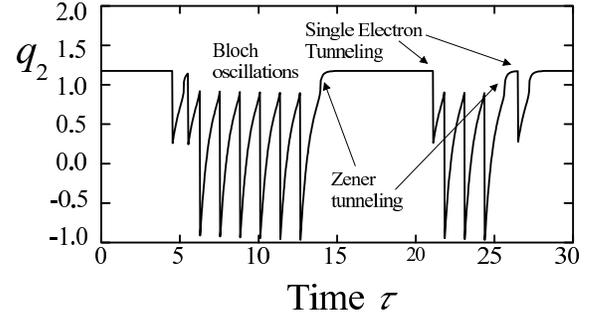}
\caption{ The quasicharge $q_{2}\,$versus time at a bias point denoted by a
white circle in Fig. 3. The Bloch oscillations are interrupted by Zener
tunneling, which drives the junction to the first energy band.\ A single
electron tunneling through junction \#1 from the base recovers the system to
the zeroth band.}%
\end{center}
\end{figure}
%EndExpansion

Figure 3 shows the normalized collector current $i_{C}$ as a function of the
bias voltage $v_{CC}$ for different base voltages $v_{B}$. The normalized bias
resistance $r=0.5$. For bias voltages $v_{CC}\gtrsim0.5$ the potential becomes
high enough so that Cooper pairs are injected through the Josephson junction.
In that range the system Bloch oscillates. Owing to the Zener tunneling the
system tends to go the upper energy band and stay there. A single electron
tunneling through the junction \#1 to the base recovers the system to the
lower energy band and the system starts to Bloch oscillate again. This is
illustrated in Fig. 4 in the bias point shown in Fig. 3 as an open circle. The
share of time the system Bloch oscillates is controlled by the base current.
The collector current is thus determined by the base voltage and it is less
than the maximum $v_{CC}/r$. Note that in practice $\left|  v_{CC}\right|
\lesssim2\Delta^{\prime}$ and $\left|  v_{CC}-v_{B}\right|  \gtrsim
\Delta^{\prime}$ are not satisfied for all $v_{CC},v_{B}$, and thus the
transistor cannot be biased at all points of operation shown in Figs. 3. The
dimensionless energy gap is $\Delta^{\prime}$.%

%TCIMACRO{\FRAME{ftbpFU}{224.75pt}{362.6875pt}{0pt}{\Qcb{The normalized
%collector current $i_{C}\,$and base current $i_{B}$ as a function of the
%normalized base voltage $v_{B}$ at $v_{CC}=1.7$ for different load resistances
%$r$. The inset shows the collector current as a function of the base current
%for $r$ varying from $1.5$ to $3.5$.}}{}{Figure}%
%maintain-aspect-ratio TRUE;  display "USEDEF";  valid_file "T";
%width 224.75pt;  height 362.6875pt;  depth 0pt;  original-width 225.25pt;
%original-height 364.75pt;  cropleft "0";  croptop "1";  cropright "1";
%cropbottom "0";  tempfilename 'HES0AS04.wmf';tempfile-properties "XPR";}}}%
%BeginExpansion
\begin{figure}
[ptb]
\begin{center}
\includegraphics[width=7cm]{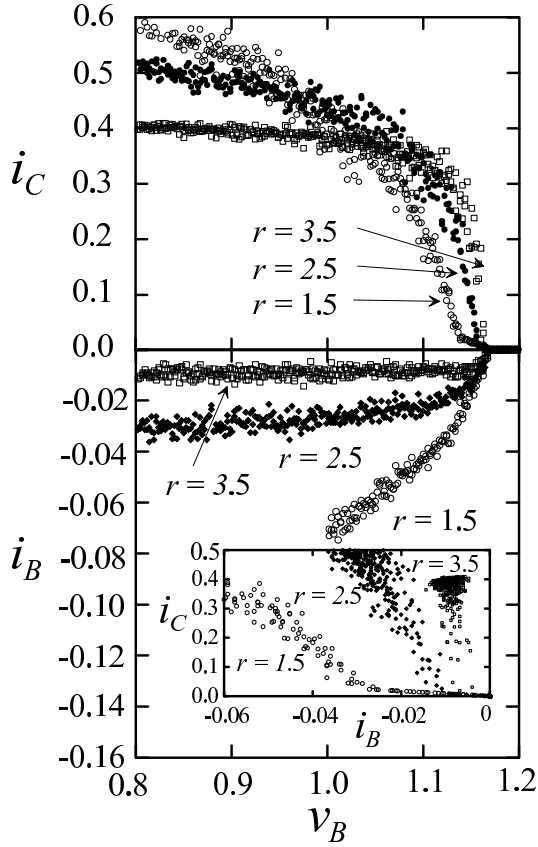}
\caption{The normalized collector current $i_{C}\,$and base current $i_{B}$ as
a function of the normalized base voltage $v_{B}$ at $v_{CC}=1.7$ for
different load resistances $r$. The inset shows the collector current as a
function of the base current for $r$ varying from $1.5$ to $3.5$.}%
\end{center}
\end{figure}
%EndExpansion

Figure 5 shows the collector and base currents as a function of the base
voltage. The input impedance $r_{in}=dv_{B}/di_{B}$ varies from $1$ to very
high values when $r$ is varied from $1.5$ to $3.5$. The normalized
transconductance gain $g^{\prime}{}_{m}=di_{C}/dv_{B}$ tends to increase with
increasing bias resistance: the maximum $g^{\prime}{}_{m}$ is about $6$ for
$r=1.5$ and close to $15$ for $r=3.5$. The maximum current gain $\beta
^{\prime}=di_{C}/di_{B}=g^{\prime}{}_{m}r_{in}$ obtained for $r=1.5$ is about
$10$ but when $r=3.5$ the current gain becomes arbitrarily high as shown in
the inset. In SI units $R_{in}$ varies from $R_{T1}$ to arbitrarily high
values, the maximum transconductance gain $g_{m}$ varies from $6/R_{T1}$ to
$15/R_{T1}$. Owing to the high input impedance and high current gain the sj
BOT can also provide power gain.

In this paper we have described the basic operating principle of the BOT. A
single junction BOT where the system is isolated with the bias resistance was
analyzed in detail. We were able to show that the Bloch oscillations can be
controlled by single electrons tunneling in such a way that the system
provides high current gain. We have also analyzed the sj BOT where the charge
is injected to the island via an $LR$-circuit.\cite{ascbot} It is also
possible that a BOT can be realized by replacing the bias resistance with an
other mesoscopic junction. We call such a device a double junction BOT. The
noise of the BOT at a finite temperature is limited by the thermal noise of
the tunneling resistance $R_{T1}$ and the shot noise related to the current
trough the junction \#1. According to a brief analysis the noise temperature
of the BOT remains below the bath path temperature and the optimal impedance
exceeds $R_{T1}$.\cite{ascbot} Since the optimal source resistance is
reasonably low at least compared to that of the SET, the BOT may be used as a
preamplifier for detectors. An other application for the BOT could be an
experiment where a voltage drop across a resistance induced by a current from
a single electron pump is compared against a Josephson voltage standard.

\end{document}